\documentstyle[12pt]{article}
\begin{document}
\addtolength{\baselineskip}{0.4\baselineskip}
\thispagestyle{empty}
\rightline{SNUTP-94-73}
\rightline{CALT-68-1944}
\vskip 1.5cm
\centerline{\bf TOPOLOGY OF A NONTOPOLOGICAL MAGNETIC MONOPOLE}
\vskip 1cm
\centerline{Choonkyu Lee}
\vskip 2mm
\centerline{\it Department of Physics  and Center for Theoretical Physics}
\centerline{\it Seoul National University, Seoul 151-742, Korea}
\vskip 7mm
\centerline{Piljin Yi\footnote{e-mail address: piljin@theory.caltech.edu}
\footnote{Address after Sept. 1, 1994: \it Physics Department,
Columbia University, New York, NY 10027, U.S.A.}}
\vskip 2mm
\centerline{\it California Institute of Technology, Pasadena,
CA 91125, U.S.A.}

\vskip 2cm
\centerline{ABSTRACT}
\begin{quote}
Certain nontopological magnetic monopoles, recently found by Lee and
Weinberg, are reinterpreted as topological solitons of a non-Abelian
gauged Higgs model. Our study makes the nature of the Lee-Weinberg
monopoles more transparent, especially with regard to their singularity
structure.

\end{quote}
\vskip 3cm
\centerline{\it Submitted to Physics Letters B.}

\newpage
When Dirac \cite{Dirac} founded the theory of magnetic monopoles in 1931,
the monopole was not something that people could not live without. Things
changed a great deal in the seventies when  't Hooft and Polyakov
\cite{hp} showed that magnetic monopoles inevitably occur as solitons of
spontaneously broken non-Abelian gauge theories; such as all grand unified
theories where an internal semi-simple gauge symmetry is spontaneously
broken to $U(1)$. Their existence is understood in terms of the nontrivial
topology of the vacuum manifold, and as such the non-Abelian nature of
the original gauge group plays a crucial role. In particular, the Dirac
quantization rule is naturally enforced by the underlying non-Abelian
structure.

Recently, however, Lee and Weinberg \cite{LW} constructed a new class of
finite-energy magnetic monopoles in the context of a purely Abelian
gauge theory. Amazingly enough, the corresponding $U(1)$ potential is
simply that of a point Dirac monopole (with the monopole strength satisfying
the Dirac quantization rule), yet the total energy is rendered
finite by introducing a charged vector field of positive gyromagnetic
ratio and by fine-tuning a quartic self-interaction thereof. For more
general values of the couplings, this theory together with Einstein gravity
was found to produce new magnetically charged black hole solutions
with hair \cite{LW}.

One might be tempted to conclude from this that the existence of the
magnetic monopoles does not require an underlying non-Abelian structure,
let alone a nontrivial topology of the vacuum manifold. We believe
this is a bit premature, and is misleading as far as this particular
model is concerned. As pointed out by Lee and Weinberg \cite{LW}, their
$U(1)$ theory may be regarded as a gauge-fixed version of the usual $SO(3)$
Higgs model at some special values of the couplings. An important question
to ask here is whether there exists such a hidden structure at
other values of the couplings as well. In this letter, we will show that
there is indeed a hidden non-Abelian gauge symmetry for general values
of the couplings, and that subsequently the {\it integer-charged}
monopoles of Lee and Weinberg may be regarded as topological solitons
associated with certain nonrenormalizable deformations of the $SO(3)$ Higgs
model. Adopting the radial gauge rather than the unitary gauge, we found
that the apparent Dirac string disappears as usual, while the singularity
at the origin still needs to be examined. An important byproduct of our
study is a topological understanding of the Dirac quantization rule
for the integer-charged Lee-Weinberg monopoles.

The Abelian model of Ref.\cite{LW} consists of a $U(1)$ electromagnetic
potential $A_\mu$, a charged vector field $W_\mu$, and a real scalar $\phi$.
The Lagrange density was chosen to have the form
\begin{eqnarray}
{\cal L}&=&-\frac{1}{4}F_{\mu\nu}F^{\mu\nu} -\frac{1}{2}\,\vert \bar{D}_\mu
           W_\nu-\bar{D}_\nu W_\mu \vert^2 + \frac{g}{4}H_{\mu\nu}F^{\mu\nu}
           -\frac{\lambda}{4}H_{\mu\nu}H^{\mu\nu} \nonumber \\
        & &-\frac{1}{2}\,\partial_\mu \phi\,\partial^\mu\phi
           -m^2(\phi)\,\vert W_\mu \vert^2 -V(\phi), \label{eq:LLW}
\end{eqnarray}
where $F_{\mu\nu}=\partial_\mu A_\nu -\partial_\nu A_\mu$ and $\bar{D}_\mu
W_\nu= \partial_\mu W_\nu+ie\,A_\mu W_\nu$. The magnetic moment
$H_{\mu\nu}$ is given by the antisymmetric product $ie\,(W^*_\mu W_\nu-
W^*_\nu W_\mu)$ so that $g$ is the gyromagnetic ratio of the charged vector.
Here, $g$ and $\lambda$ are positive constants, while $m(\phi)$ vanishes
at $\phi=0$ but is equal to $m_W\neq 0$ when $\phi$ is at its (nontrivial)
vacuum value. As mentioned above, when $g=2$, $\lambda=1$ and
$m(\phi)= e\phi$, this is nothing but the unitary gauge version of the
spontaneously broken $su(2)$ gauge theory of Ref.\cite{hp}, and thus
renormalizable. But for generic values of $g$ and $\lambda$, the theory
is nonrenormalizable.

The ansatz for the unit-charged, spherically symmetric, Lee-Weinberg
monopole can be written most succinctly in term of the differential forms
$A\equiv A_\mu \,dx^\mu$ and $W\equiv W_\mu\, dx^\mu$:
\begin{equation}
\phi=\phi(r), \qquad A= \frac{1}{e}\,(\cos\theta-1)\,d\varphi , \qquad
W= \frac{i\,u(r)}{e\sqrt{2}}\,[\,\exp{i\varphi}\,][\,d\theta +i\sin\theta
\,d\varphi\,]. \label{eq:ansatz}
\end{equation}
$u$ and $\phi$ are radial fuctions to be determined by the field equations
and appropriate boundary conditions.
Note that the electromagnetic potential is simply that of a point-like
Dirac monopole and the field configuration appears singular at origin,
in addition to the presumably harmless Dirac string along the negative
$z$-axis. A key observation in Ref.\cite{LW} is that, if the relationship
$4\lambda=g^2$ holds, the total energy of the resulting solution can
be actually finite despite this singular behaviour at origin. In addition
to such a unit-charged solution, the Dirac quantization rule allows all
integral and half-integral magnetic charges.

Let us turn to the question of the hidden non-Abelian structure.
Consider the following nonrenormalizable $SO(3)$ gauge theory with
the gauge connection 1-form $B=(B^a_\mu \,dx^\mu)\, T_a$ and a triplet Higgs
$\Phi =\Phi^a T_a$.
\begin{eqnarray}
{\cal L}'&=& -\frac{1}{4} G_{\mu\nu}^a G^{a\mu\nu}
             +\frac{g-2}{4}{\cal F}_{\mu\nu}{\cal H}^{\mu\nu}
             -\frac{\lambda-1}{4}{\cal H}_{\mu\nu}{\cal H}^{\mu\nu}
             \nonumber \\
         & & -\frac{1}{2}(D_\mu \Phi)^a(D^\mu \Phi)^a
             -\frac{1}{e^2}(m^2(|\Phi|)-e^2\Phi^a\Phi^a)\,(D_\mu \hat{\Phi})^a
              (D^\mu \hat{\Phi})^a -V(|\Phi|)
\end{eqnarray}
$G$ is the non-Abelian gauge field strength associated with $B$, while
$D_\mu \Phi^a\equiv\partial_\mu \Phi^a+\epsilon^{abc}B^b_\mu \Phi^c$
and $\hat{\Phi}^a\equiv \Phi^a/|\Phi|$. In addition, we defined two
gauge-invariant antisymmetric tensor ${\cal F}$ and ${\cal H}$
as follows, \footnote{$\cal F$ was previously used by `t Hooft \cite{hp}
to represent physical electromagnetic fields.}
\begin{equation}
{\cal F}_{\mu\nu}-{\cal H}_{\mu\nu}\equiv G^a_{\mu\nu}\hat{\Phi}^a,
\qquad {\cal H}_{\mu\nu}\equiv -\frac{1}{e}\,\epsilon_{abc} \hat{\Phi}^a
(D_\mu \hat{\Phi})^b (D_\nu \hat{\Phi})^c .
\end{equation}
Now in the unitary gauge $\hat{\Phi}^a=\delta^{3\,a}$, we may identify
$\phi$ with $|\Phi|$, $A_\mu$ with $B^3_\mu$, and $W_\mu$ with $(B^1_\mu+
iB^2_\mu)/\sqrt{2}$. This results in the following reduction formulae:
\begin{eqnarray}
{\cal F}_{\mu\nu} &\Rightarrow & F_{\mu\nu}, \nonumber \\
{\cal H}_{\mu\nu} &\Rightarrow & H_{\mu\nu},\nonumber \\
G^3_{\mu\nu} &\Rightarrow & F_{\mu\nu}-H_{\mu\nu}, \nonumber \\
G^1_{\mu\nu}+i G^2_{\mu\nu} &\Rightarrow & \sqrt{2}
\,(\bar{D}_\mu W_\nu -\bar{D}_\nu W_\mu), \nonumber \\
(D_\mu \hat{\Phi})^a(D^\mu \hat{\Phi})^a &\Rightarrow & 2e^2\, W^*_\mu W^\mu,
\nonumber \\
(D_\mu {\Phi})^a(D^\mu {\Phi})^a &\Rightarrow & \partial^\mu \phi \,
\partial_\mu \phi +2e^2\phi^2\, W^*_\mu W^\mu \nonumber.
\end{eqnarray}
Expanding the $G_{\mu\nu}^a G^{a\mu\nu}$ term, one easily notices that
the $SO(3)$ invariant Lagrangian $\cal L'$ reduces to $\cal L$ of the
apparently Abelian theory of Lee and Weinberg.

According to the standard argument \cite{Coleman}, all configurations with
a regular and finite asymptotic behaviour may be classified in terms of a
topological winding number:
\begin{equation}
n=\frac{1}{8\pi}\oint dS^i\epsilon_{ijk}\epsilon_{abc}\hat{\Phi}^a
\partial_j \hat{\Phi}^b \partial_k \hat{\Phi}^c , \label{eq:winding}
\end{equation}
where the integral is over the asymptotic two-sphere. Note that $n$ is an
integer in general, and the magnetic monopole strength is related to this
number by $g_{magnetic}=-4\pi n/e$ \cite{Ara}. The unit-charged case ($n=1$)
is of special interest, because the corresponding solution has the spherical
symmetry. For example, if $g=2$, $\lambda=1$ and $m=e\phi$, the $n=1$
solution is given by the usual 't Hooft-Polyakov monopole.

Actually, the ansatz (\ref{eq:ansatz}) for the sperically symmetric
Lee-Weinberg monopole is exactly the same as that of the 't Hooft-Polyakov
monopole but written in the unitary gauge. Written in the so-called radial
gauge, the ansatz for the 't Hooft-Polyakov monopole has the following
well-known hedgehog configuration:
\begin{equation}
\Phi^a= \hat{x}^a\,\phi(r), \qquad B^a=-\frac{\epsilon_{abc}\,
x^b \, dx^c}{e\,r^2}\,[\,1-u(r)\,] . \label{eq:hedge}
\end{equation}
Gauge transforming to the unitary gauge,
\begin{equation}
\Omega^{-1}\, \hat{x}^aT_a\, \Omega = T_3 \qquad \hbox{with}
\quad \Omega \equiv e^{-i\varphi T_3}e^{-i\theta T_2}e^{i\varphi T_3},
\end{equation}
a lengthy but straightforward calculation shows that the hedgehog
configuration (\ref{eq:hedge}) is indeed  gauge-equivalent to the ansatz
(\ref{eq:ansatz}). This clearly demonstrates that the spherically symmetric
unit-charged Lee-Weinberg monopole has the topological winding number $n=1$,
and that the Dirac string apparent in (\ref{eq:ansatz}) is superfluous.
Similar procedures for $n>1$ non-Abelian configurations should also lead
to all integer-charged Lee-Weinberg monopoles automatically,\footnote{What
about half-integer-charged monopoles of Lee and Weinberg \cite{LW}? The
topological quantization (\ref{eq:winding}) clearly dictates that it is
not possible to remove the Dirac string of such $U(1)$ monopoles by lifting
the configuration to the underlying $SO(3)$ and by performing a non-Abelian
gauge transformation. Instead, the lifted Dirac string will be carrying a
$Z_2$ flux that is again undetectable. Here, $Z_2$ corresponds to the
fundamental group of $SO(3)$.} although, due to the lack of the spherical
symmetry, the explicit forms of the necessary gauge transformations are more
difficult to find.

\vskip 5mm
Now that the Dirac string disappeared by virtue of the radial gauge, let us
concentrate on the possible singular structure at origin $r=0$. Generalizing
to include dyons \cite{Zee} while maintaining $n=1$, we may modify the
hedgehog configuration (\ref{eq:hedge}) to have nontrivial $B_t^a$.
\begin{equation}
\Phi^a= \hat{x}^a\,\phi(r), \qquad B^a=-\frac{\epsilon_{abc}\,
x^b \, dx^c}{e\,r^2}\,[\,1-u(r)\,]-\hat{x}^a v(r)\,dt . \label{eq:hedge2}
\end{equation}
On such configurations, the total energy is given by the following
functional of $u$, $\phi$, and $v$.
\begin{eqnarray}
{\cal E}=\int dx^3 \hskip -2mm&\biggl\{&\hskip -2mm \frac{1}{e^2r^2}\,
\biggl(\frac{du}{dr}\biggr)^2+\frac{1}{2}\,\biggl(\frac{dv}{dr}\biggr)^2+
\frac{1}{2}\biggl(\frac{d\phi}{dr}\biggr)^2 \nonumber \\
& &\hskip -6mm +\frac{1}{2e^2r^4}\,[\,\lambda u^4-g u^2+1\,]+ \frac{u^2}{e^2
r^2}\,[\,m^2+e^2v^2\,] +V(\phi) \hskip 1.5mm \biggr\}. \label{eq:E}
\end{eqnarray}
For the integrand, we evaluated $T_{00}$, the energy-density associated with
$\cal L'$, on the configuration (\ref{eq:hedge2}) using such radial gauge
results as \ ${\cal F}_{ij}=-\epsilon_{ija}x^a/er^3$ and ${\cal H}_{ij}=
u(r)^2 {\cal F}_{ij}$. Clearly, for the configuration (\ref{eq:hedge2})
to be regular at origin, we need to have $u(0)=1$, $v(0)=0$, and $\phi(0)=0$,
while, for a finite total energy, we must also require
\begin{equation}
\lambda \,u^4(0)-g \,u^2(0)+1=0 .\label{eq:finite}
\end{equation}
Only if this last condition is met, the integrand of the energy functional
$\cal E$ does not have any $r^{-4}$ singularity near origin,

Does there exist an everywhere regular finite energy solution of
the form (\ref{eq:hedge2})? The answer is no in general, such a solution being
possible only when $g=2$ and $\lambda=1$. This can be easily seen as follows.
To be a static solution to the field equations, the configuration must
extremize the energy-functional. Suppose $\bar{u}(r)$ is such a regular
finite energy solution, and consider a small variation $\delta u(r)$
around it such that $\delta u(0)=0$. Varying $\cal E$, we now observe that
the only contributions that may lead to a small-$r$ singularity of $r^{-3}$
are given by the following variation,
\begin{equation}
{\cal A}\equiv\int dx^3 \frac{1}{2e^2r^4}\,[\,\lambda u^4-g u^2+1\,] \quad
\Rightarrow \quad \delta{\cal A}=\int dx^3 \frac{1}{2e^2 r^4}\,
[\,4\lambda u^3-2g u\,]\,\delta u+\cdots . \label{eq:A}
\end{equation}
Choosing $\delta u(r)=r+O(r^2)$, we find $4\lambda\,\bar{u}^3(0)-2g\,
\bar{u}(0)=4\lambda-2g=0$ as a necessary condition for $\bar{u}(r)$
to satisfy the equation of motion. But the finite energy
condition (\ref{eq:finite}) requires $g=h+1$ for such a regular solution,
which is not compatible with $4\lambda-2g=0$ unless $g=2$ and $\lambda=1$.
Thus the existence of a regular finite-energy monopole solution demands
the special values of vector couplings, $g=2$ and $\lambda=1$, appropriate
for the renormalizable field theory.

However, we do obtain a finite-energy solution if we choose not to insist on
the everywhere regularity of the configuration (\ref{eq:hedge2}). The
finite-energy monopole of Lee and Weinberg belongs to this class. Instead
of demanding $u(0)=1$, let $u(r)$ approach a finite constant $C$ as
$r\rightarrow 0$ but still let $v(0)=\phi(0)=0$. With such a singular
boundary condition, the finite-energy requirement and the stationary argument
above translate into the following two conditions:
\begin{equation}
\lambda\,C^4-g\,C^2+1=0, \qquad 4\lambda \,C^3-2g\,C=0.
\end{equation}
This in turn leads to the Lee-Weinberg condition $4\lambda =g^2$ with $g>0$
as well as to the boundary condition $u^2(0)={g/2\lambda}=2/g$ \cite{LW}.
More careful study of such a singular finite-energy monopole may be carried
out with the help of the following equations of motion for these radial
functions,
\begin{eqnarray}
& &\frac{d^2u}{dr^2}+(\frac{g}{2}-\lambda u^2)\,\frac{u}{r^2}=
(m^2+e^2 v^2)\,u \nonumber \\
& &\frac{d^2\phi}{dr^2}+\frac{2}{r}\,\frac{d\phi}{dr}-\frac{u^2}{e^2r^2}\,
\frac{d\,m^2}{d\phi} =\frac{dV}{d\phi} \nonumber \\
& &\frac{d^2 v}{dr^2}+\frac{2}{r}\,\frac{dv}{dr}- \frac{2u^2}{r^2}\,v=0
\nonumber
\end{eqnarray}
For instance, a linearization of this nonlinear system near origin predicts
that the solutions are in general non-analytic at $r=0$. The same technique
reproduces the well-known analytic behaviour, $u(r)=1+O(r^2)$ and
$\phi=1+O(r)$,  of the 't Hooft-Polyakov monopoles when the
couplings are those of the renormalizable theory.

The condition $4\lambda =g^2$ that is necessary for a finite-energy solution
to exist can be also recognized on the basis of the energy-functional $\cal E$
itself. The only part of $\cal E$ that is not necessarily positive definite
is the quantity $\cal A$ of (\ref{eq:A}). But the integrand of $\cal A$ is
quadratic in $u^2(r)$ with coefficients $\lambda$ and $g$, and subsequently
the energy functional $\cal E$ is bounded from below if and only if $4\lambda
\ge g^2$. In particular when $4\lambda=g^2$, it is easy to see that there
exists a finite energy configuration of type (\ref{eq:hedge2}) (but not
necessarily a solution yet). In that case, we may secure a local minimum of
$\cal E$, thus a finite-energy {\it solution}, by suitably adjusting $u(r)$,
$\phi(r)$ and $v(r)$.  If $ 4\lambda >g^2$ ($4\lambda <g^2$) instead,
monopole solutions are of infinite positive (negative) energy, but
nevertheless physically significant if coupled to gravity \cite{LW}.

\vskip 5mm
In this letter, we presented a new interpretation of the integer-charged
Lee-Weinberg monopoles in terms of a topological structure arising in a
nonrenormalizable extension of the 't Hooft-Polyakov monopole model.
In the unitary gauge where the topological nature is obscured, the
monopoles are riddled with Dirac strings, albeit harmless,
and the singularity structures of the underlying non-Abelian fields
are less transparent. We reexamined the core of the unit-charged monopole
from the radial gauge viewpoint and reproduced some of the results in
Ref.\cite{LW}.

An obvious follow-up question to ask\footnote{We are grateful to E.
Weinberg for suggesting this.} is: would it always be possible to
provide a non-Abelian gauge theoretic interpretation for Lee-Weinberg-type
monopoles, when the theory involves several charged vector fields? We find
this unlikely. For example, consider the following generalization of
Lee-Weinberg model with the $N$ charged vector fields $W^{(n)}, \; n =1,
\ldots, N$ and the corresponding magnetic moment tensors $H^{(n)}_{\mu\nu}$:
\begin{eqnarray}
{\cal L}_{N}&=&-\frac{1}{4}F_{\mu\nu}F^{\mu\nu} -\frac{1}{2}\sum_n \vert
               \bar{D}_\mu W^{(n)}_\nu-\bar{D}_\nu W^{(n)}_\mu \vert^2
               \nonumber \\
            & &+\frac{1}{4}\sum_n g_n H^{(n)}_{\mu\nu}F^{\mu\nu} -\frac{1}{4}
               \sum_{n,m} \lambda_{nm} H_{\mu\nu}^{(n)} H^{(m)\mu\nu}+\cdots.
\end{eqnarray}
As long as the relationship $4\lambda_{nm}=g_n g_m$ holds, an ansatz,
entirely analogous to (\ref{eq:ansatz}) of the $N=1$ case, leads to a
spherically symmetric finite-energy monopole with the boundary condition
satisfying $\sum_{n=1}^N \,g_n\,[u^{(n)}(0)]^2=2$. But it is rather difficult
to imagine why there should exist non-Abelian interpretations for all such
generalized theories, specifically for all $N$ and for all values of the vector
couplings. On the other hand, what one may expect to be true is that
the resulting nontopolgical monopoles are generically singular at
origin despite the finite total energy and that, just as we have observed in
$N=1$ case above, the complete regularity is recovered only when the
relevant vector coupling structure is renormalizable, possibly corresponding
to a spontaneously broken non-Abelian gauge theory.

\vskip 1cm
\centerline{\bf Acknowledgment}
\vskip 5mm
We are very grateful to K. Lee and E. Weinberg for sharing their insight and
also for critical comments on this work. C.L. enjoyed useful conversations
with V. P. Nair, and P.Y. would like to thank H-K Lo for an illuminating
discussion on the topological charge and the Dirac quantization.
The work of C.L. is supported in part by the Korean Science and Engineering
Foundation (through the Center for Theoretical Physics at SNU) and the
Ministry of Education, Republic of Korea. The work of P.Y. is supported
in part by D.O.E. Grant No. DE-FG03-92-ER40701.

\end{document}